\documentstyle[epsf,12pt]{article}
\begin{document}
\begin{center}
\Large{\bf Can Recent Charge Fluctuation Calculations Be a Reliable
Signal for a QGP at RHIC?}\\
\large{S.J. Lindenbaum$^{a,b}$ and R.S. Longacre$^a$\\
$^a$Brookhaven National Laboratory, Upton, NY 11973, USA\\
$^b$City College of New York, NY  10031, USA\footnote{This research was 
supported by the U.S. Department of Energy under Contract No. 
DE-AC02-98CH10886 and the City College of New York Physics Department}}
\end{center}

\begin{abstract}
The recent papers of Jeon and Koch\cite{jeon} and Asakawa, Heinz and 
Muller\cite{asakawa} propose that the event by event fluctuations of the 
ratio of the positively charged and negatively charged pions could provide 
a distinct signal for a QGP at RHIC/LHC due to differences in those from 
the QGP phase, and the Hadron Gas Phase. In this paper we point out that 
aside from the questionability of the many assumptions in the treatment 
used, even following their approach there are other effects not considered,
e.g. color charge fluctuations which we show could signifiantly or 
completely wash out the proposed signal. Therefore lack of observation of 
these charge flucuation signals cannot lead one to conclude that a QGP is 
not formed at RHIC. A general discussion of experimental requirements for 
observation of such signals (if they exist) and how to interpret them 
is included.
\end{abstract}

\section{Introduction}

The recent papers of Jeon and Koch\cite{jeon} and Asakawa, Heinz, and
Muller\cite{asakawa} argue that the event by event fluctuations of the 
ratio of the positively charged and negatively charged pions provide a
distinct signal for Quark-Gluon Plasma at RHIC and eventually at LHC.
Since the size of the average fluctuations of electric charge differ 
widely in the confined and deconfined phases, it is possible that these 
initial state fluctuations survive until freezeout and thus appear 
differently in the final state for pions arising from a QGP and those 
from normally confined processes.

This is a complex process and thus their novel work involves many
assumptions and approximations which one can question whether they lead to 
a reliable conclusion. However, a second perhaps more relevant question is
following their approach, are there effects they have not considered 
which may significantly or even totally wash out the predicted signals?
In this paper we point out that fluctuations in color and anti-color if
taken into account may significantly or even more or less totally wash
out the predicted effects.

\section{Calculation of Charge Fluctuations Following \cite{jeon} and 
\cite{asakawa}}

We now follow the arguments made in \cite{jeon} and \cite{asakawa} and then
consider the case with the addition of our arguments.

If as discussed one can assume that conditions are such that the expansion 
is too fast for local fluctuations to follow the mean thermodynamic 
evolution of the system, and concern ourselves with locally conserved 
quantities that show large differences in a hadron gas (HG), and a 
Quark-Gluon Plasma (QGP), the different fluctuations in these quantities 
may survive the freezout.  If sub-volumes are considered which move rapidly 
away from one another, conditions which one expects, due to a strong 
differential flow pattern\cite{ackerman}, and thus these fluctuations 
would reflect the phase, HG or QGP from which the pions came from.
At RHIC, mesons dominate the hadronic particles, and the baryonic chemical
potential $\mu$ is much less than the temperature. The final state 
hadrons are mostly pions, although approximately 30-50\% of these pions 
are estimated to come from resonance decay in calculations in [\cite{jeon} 
and \cite{asakawa} respectively.

Let us now consider charge fluctuations of pions in an HG compared to a 
QGP: Both \cite{jeon} and \cite{asakawa} use ideal gases in equilibrium 
for their estimates.  However, \cite{jeon} also uses the lattice 
calculations of \cite{gottlieb} to determine $D$ as approximately 1 for a 
QGP, and \cite{asakawa} obtains a similar effect.  Following \cite{jeon} 
and \cite{asakawa} we will use the ideal gas approximation in our 
considerations. Take a phase space sub-volume which has $N$ hadrons. Let 
us assume that we have only light quark mesons and meson resonances which 
decay. Jeon and Koch define a parameter $D =<N_{ch}><\delta R^2>$ which 
measures the mean charge multiplicity times the fluctuation of the ratio 
$R$, where
\begin{equation}
R={N_+\over N_-}.
\end{equation}

$N_+$ and $N_-$ are the positive and negative particle (pion) multiplicities
The fluctuation of $R$ is shown to be
\begin{equation}
<\delta R^2>= <R^2> - <R>^2.
\end{equation}

>From \cite{jeon}, we have
\begin{equation}
D=<N_{ch}><\delta R^2>.
\end{equation}

where

\begin{equation}
<N_{ch}>=N_+ + N_-.
\end{equation}

The net charge $Q$ is
\begin{equation}
Q=N_+ - N_-.
\end{equation}

Reference \cite{jeon} also shows that $<\delta Q^2>$ is purely Poisson, 
using thermal distributions and disregarding correlations, is approximately 
for a pion gas  equal to $<N_{ch}>$, thus we have
\begin{equation}
<\delta Q^2> \approx <N_{ch}>
\end{equation}

Reference 1 then shows that $D$ becomes
\begin{equation}
D\approx 4{{<\delta Q^2>}\over N_{ch}},
\end{equation}
thus
\begin{equation}
D\approx 4.
\end{equation}
         
The value of $D$ for that coming from a thermal light quark meson
and meson resonance system,  the resonances of which decay into charged
pions (for example the neutral $\rho$ which decays into a positive and
negative pion), reduces the observed $D$ to 

\begin{equation}
D\approx  3 {\rm ~from~[1],~and~perhaps~as~low~as~2~from~[2].}
\end{equation}

Therefore, $D$ is reduced because there are some neutral meson resonaces 
($\rho$, etc.) which decay into $\pi^+$ and $\pi^-$, thus decreasing the 
fluctuations. Hence for a thermal light quark resonace meson system

\begin{equation}
D = 4{{<\delta Q^2>}\over <N_{\rm meson}>} = \approx 3-2
\end{equation}

as estimated by references\cite{jeon} and \cite{asakawa} respectively.

In references \cite{jeon} and \cite{asakawa} for a QGP they both estimate 
$D$ approximately 1. Thus the $D$ signal for mesons decreases by a factor 
of 2-3  for those pions which come from a QGP.

\section{Taking into Account Color Fluctuations}

In the calculation that follows we will take into account color charge 
$<\delta Q^2_{\rm color}>$ fluctuations and, anti-color charge 
$<\delta Q^2_{\rm anti-color}>$ fluctuations, since it appears there is 
no reason these fluctuations would not occur in the sub-volume before 
passing from the QGP to the meson system. At this stage of the system the 
color charge degree of freedom is active and important. These color and 
anti-color fluctuations should be frozen out just like the charge 
fluctuations (as \cite{jeon} and \cite{asakawa} have done for electric 
charge flucuations). Their reduction is not expected to be local since, for 
example, if it's done by the movement of quarks and anti-quarks in and out 
of the sub-volume  in order to eliminate the color and anti-color charge 
fluctuations. In our subsequent treatment of color charge fluctuations, 
and the fragmentation process, we will find the $D_{QGP}$ can become 
approximately equal to the $D_{HG}$ if the fragmentation size is large 
enough to mix the sub-volumes. Thus the signal for the QGP could be washed 
out. 

As we have stated, this problem is a difficult and very complex one, and we
do not claim to have solved it. However the questions regarding color
charge fluctuations, etc. that we are raising should be addressed, and a
critical evaluation of the reliability of fluctuation calculations made.

\section{A Model of Charge Fluctuations Including Color Charge Fluctuations
and Hadronization}

The following model is not claimed to be unique or even a reasonable
representation of the real physical situation. Its purpose is merely
to show that color charge fluctuations could drastically affect the
charge fluctuation calculations of \cite{jeon} and \cite{asakawa}. In 
addition we beleive the physical situation can be much more complex than 
any of the models treating it.

Let us consider the quark gluon plasma, following the approach of \cite{jeon}
and \cite{asakawa}.  The inital number of quarks, anti-quarks, and gluons is 
equal to the number of mesons after hadronization, assuming conservation of
entropy\cite{jeon, asakawa}. For a given sub-volume $<\delta Q^2>$ is 
fixed and will not change in the final hadronization of the quark qluon 
plasma.  Thus one would expect a $<\delta Q^2>$ for the QGP. 
Reference \cite{jeon} calculated $D$ to be a factor of approximately 3 
smaller than for a hadron gas, while Ref.\cite{asakawa} also claimed a 
similar though smaller (factor of $\approx 2$) result using the same 
conserved $<\delta Q^2>$ when the QGP hadronizes.

If we redefine equations (7 and 10), it becomes easier to keep track of the 
steps in hadronization in the system.  We generalize $D$ as
                             
\begin{equation}
D = 4{{<\delta Q^2>}\over <N_{\rm particles}>} 
\end{equation}

With this new definition of $D$, entropy conservation which is assumed
will not mean that $D$ stays the same as in  Ref.[1] which shows that for a
QGP, $D$ is approximately 1 and that $<\delta Q^2> \approx {1\over 4}
<N_{\rm particles}>$ where $N_{\rm particles}$ is the number of quarks,
anti-quarks and gluons. Thus in moving from the QGP to the meson system
$<\delta Q^2>$ remains the same and the number of mesons are equal
to the number of quarks, anti-quarks and gluons, therefore our generalized 
$D$ is unchanged like the old $D$ of reference \cite{jeon}.

We will now reconsider our classic meson light quark system. If we 
recalculate (10) using the quarks and anti-quarks which make up the mesons 
$D$ becomes equal to 1.5, because $<\delta Q^2>$ remained the same and 
$N_{\rm particles}$ doubled.

In the hadronization process of making  mesons the gluons in the QGP, 
which contain $\sim {1\over 2}$ of the degrees of freedom, in our model we
will create quark and anti-quark pairs, which  would increase the number 
of quarks and anti-quarks to double the total number. We believe that this 
is the final stage in which $<\delta Q^2>$ is conserved in any hadronization 
process. Our new $D$ has dropped by a factor two, but entropy has not changed.
This is because there is a one to one mapping of the degrees of freedom 
from quark anit-quark color-singlet pairs to mesons. However additional 
quarks and anti-quarks have to be formed by the gluons in the fragmentation 
of the QGP. In the intial QGP before hadronization the quarks, anti-quarks 
and gluons are equal to the number of final mesons. The sub-volume
also has color charge fluctuations $<\delta Q^2_{\rm color}>$ and
$<\delta Q^2_{\rm anti-color}>$. At this stage the color charge degree of
freedom is active and very important. In passing from QGP to the meson 
system these fluctuations get frozen out and become part of the process 
leading to the increased number of quarks and anti-quarks just before 
forming the mesons. In this way entropy is conserved through the 
hadronization process. The reduction of $<\delta Q^2_{\rm color}>$ and 
$<\delta Q^2_{\rm anti-color}>$ is achieved through gluon fragmentation
into quark, anti-quarks and gluons.  This fragmentation is not local 
because quarks and anti-quarks move in and out of the sub-volume in order 
to get rid of the color and anti-color charge fluctuations. When the gluons
fragment and before the movement and rearrangement of color charge, the 
quarks and anti-quarks are expected to be approximately doubled. Since the 
$<\delta Q^2>$ is the same as its value in the QGP, the value of $D$ drops 
from  $\approx 1.0$ to approximately 0.5 because $N$ is approximately equal 
to double the initial particles.  However the movement of quarks and 
anti-quarks into and out of the sub-volume, which reduces the 
$<\delta Q^2_{\rm color}>$ and $<\delta Q^2_{\rm anti-color}>$ will
increase the electric charge fluctuation. How effective the color charge 
fluctuations are at increasing $<\delta Q^2>$ depends on the size of the 
gluon fragmentation region. We have studied how the size of gluon 
fragmentation increases the random mixing of quarks and antiquarks from 
zero size where color fluctuations are removed by soft gluon teleportation, 
to larger sizes where color singlets are formed by rearrangment. The maxium 
transfer of color fluctuation happens in our calculations at 3.5$fm$
gluon fragmentation size (see Fig. 1), where the value of color
fluctuations becomes equal to the value of charge fluctuations. At 
increasing $fm$ beyond 3.5 the the transfer of color fluctuation remains 
approximately the same. Thus we get a half unit increase in $D$ for the 
reduction of $<\delta Q^2_{\rm color}>$ and another half unit increase in 
$D$ for $<\delta Q^2_{\rm anti-color}>$ reduction. This causes $D$ to 
increase to the value of 1.5. The quarks and the anti-quarks form mesons 
and the value of $D$ goes from $\approx$ 1.5 to $\approx$ 3.0, (see Fig .1) 
which is the value for a light quark resonance system.\footnote{The next 
section (5) discusses the gluon fragmentation region size.} Thus in this 
approach the value of $D$ for a QGPis approximately equal to the same 
value as for pions which came from the Hadron Gas (i.e. there is no QGP 
signal).  Underlying the above discussion are the assumptions that the make 
up of the meson system from a thermal and a QGP hadronization are very 
similar. This may not be the case for strange particles and baryons 
anti-baryons. It is also assumed that fragmentation of gluons producing 
quarks and anti quarks occurs, and 
that quarks and anti-quarks rearrange to take care of color. What if gluons 
only fragment into gluons and they rearange into color-singlet glueballs, 
then there would not be the random walk of some electric charge to 
participate in the elimination of color charge fluctuations. We feel that 
this would not happen if the masses of glueballs remain large like lattice 
gauge predictions for formation in the normal vacuum. This seems to be the 
case for Glueballs calculated on the lattice since the very striking evidence 
for at least one $2^{++}$ glueball\cite{etkin} can only be explained 
by a $2^{++}$ glueball, while all alternative explanations over a period 
of almost two decades have been shown to be incorrect or not 
viable\cite{lind}. This $2^{++}$ glueball has the approximate mass of  
lattice guage calculations\cite{latt}.   If on the other hand some new medium 
rescaling of the glueball mass occurred or for some other reason the 
gluons in the QGP formed mostly glueballs then in our treatment a reduction 
in charge fluctuations could occur.

\begin{figure}
\begin{center}
\mbox{
   \epsfysize 3.0in
   \epsfbox{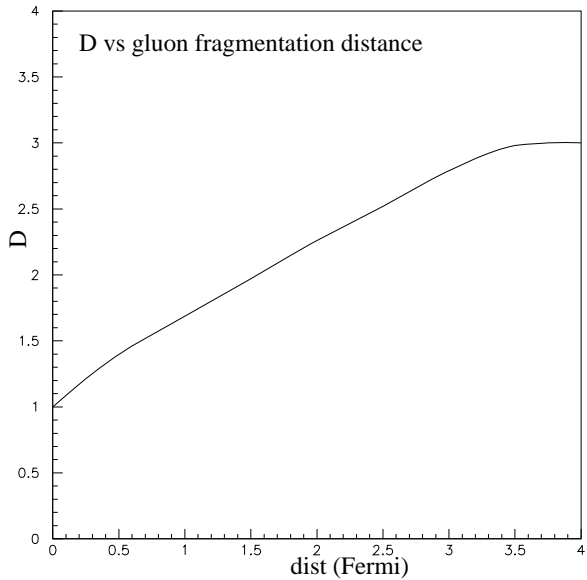}}
\end{center}
\caption{We plot $D$ for a QGP as defined in the text (Eq. 7 or 11) for 
a sub-volume versus the gluon fragmentation size dist(fermi).  The size is the 
separation distance between pairs of quarks created from gluons.}
\end{figure}

\section{Experimental Considerations}

It is clear that a measurement of charge fluctuations from a sub-volume 
of the final hadronic system is a important measurement. The experimental 
question is how can one make such a measurment? If the QGP would proceed 
through the formation of a plasma (spherical or not greatly longitudinally 
expanded) bubble as Van Hove had proposed, then even small infrequent 
suppresed local charge fluctuations would be isolated to a rapidity bump 
which would only spread out over two units. Within this signal one could  
possibly observe the charge fluctuation due to a QGP. However, if the 
bubble expands longitudinally, then the bubble could spread out to about 
six units in rapidity\cite{starhbt} and the local charge fluctuation would
increase. For relativistic heavy ion collisions of 
($\sqrt{s_{NN}} =130$) Au on Au at RHIC, produced pions at mid-rapidity and 
lower transverse momentum ($p_t$) show a source size of a radius up to 
about six $fm$\cite{starhbt}. This means that inside a given sub-volume 
pions that are moving along the beam endup in different rapidity intervals. 
Thus at a given rapidity we see low $p_t$  pions coming from different 
sub-volumes. Since the pions are a random sample of the sub-volume the 
charge fluctuation at a given rapidity would increase due to this
kinematic mixing. For the quarks and gluons used to form the light meson 
system which was used to calculate charge fluctuations shown in Fig. 1, 
we gave each parton the kinematics such that the final pions coming from 
low $p_t$ would see a six Fermi source which decreased in size as one 
increased the $p_t$. For $p_t$ above 600 MeV/c the size became 2 Fermis. 
The slope of the pion $m_t$ spectrum was 290 MeV. We then calculated 
$D$ for mid-rapity for a bin of $\Delta Y=1.0$ and $\Delta Y=2.0$ 
(see Fig. 2). In Fig. 2 we show $D$ as a function of gluon fragmentation 
size for each $\Delta Y$ bin. Even though the charge fluctuation for 
each sub-volume decreases with smaller fragmentation size, the $D$ is 
not changed much due to kinematic mixing. It is clear that the bigger 
the $\Delta Y$ bin the smaller the $D$, and how $D$ varies with $\Delta Y$
is considerably complex and model dependent.

\begin{figure}
\begin{center}
\mbox{
   \epsfysize 3.0in
   \epsfbox{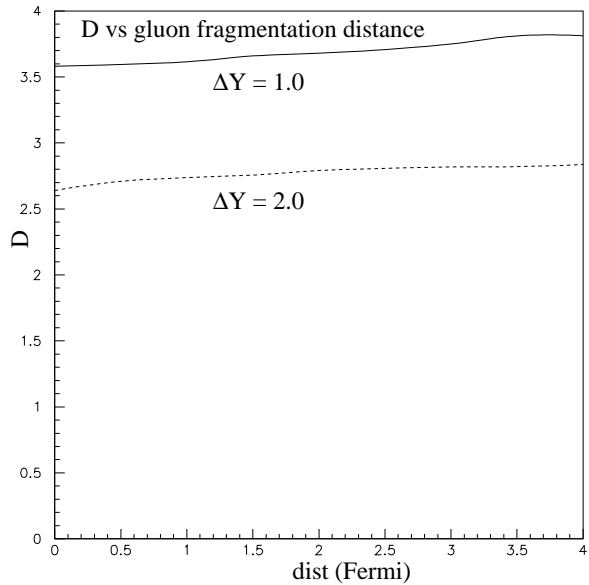}}
\end{center}
\caption{We plot $D$ for a QGP as defined in the text (Eq. 11) for particles 
lying in a $\delta Y$ range versus the gluon fragmentation size dist(fermi).
The size is the separation distance between pairs of quarks created from the 
gluons.}
\end{figure}

\section{Finding Possible Charge Fluctuations: If They Exist}

If one assumes suppressed charge fluctuations exists in some events 
containing sufficient amounts of QGP, we need a way to select those 
events and the rapidity intervals in which the plasma is contained. Large
and universal QGP production, which considering the dynamics involved as 
discussed in a recent paper\cite{lind2} is not in our opinion likely. 
Therefore some QGP selector (e.g. QGP ``Bubbles'' \cite{lind2}), is necessary 
to select an appropriate set of events. Then if there are statistically 
significant reductions of charge fluctuations in that sample that would be a 
significant dicovery.

This discovery would be consistent with some present ideas of a QGP, 
however one could not conclude that this is a unique explanation
of the effect due to the crudeness and incompleteness of the arquements 
made for it. In order to establish a QGP, numerous supporting additional 
evidence must accompany this signal. Furthermore, perhaps most importantly 
no other viable explanation of the data other than a QGP must exist.

{\it COVERSELY THE LACK OF DISCOVERY OF REDUCED CHARGE FLUCTUATIONS AT RHIC
DOES NOT IMPLY THAT QGP IS NOT FORMED AT RHIC. THE OVERLY SIMPLIFIED AND
INCOMPLETE CALCULATIONS WHICH PREDICT THIS CAN VERY EASILY BE MISLEADING
IN REGARD TO THE ACTUAL PHYSICAL SITUATION. FURTHERMORE THE QGP SIGNALS
(if they exist) MAY NOT HAVE BEEN STRONG ENOUGH TO HAVE BEEN ISOLATED 
AND OBSERVED BY THE METHODS USED.}

\section{Conclusions and Summary}

In this paper we raised the question ``Can Recent Charge Fluctuations
Calculations Be A Reliable Signal For A QGP at RHIC?''

We pointed out that the approximations and simplifications used in [1]
and [2] raises the question of their reliability  in treating this
complex process.

We then asked even following the methods of  this simple treatment, what
was left out which could change the conclusions? We pointed out that color 
charge flucuations, which were not even considered could possibly 
drastically change or even wash out the predicted effects.  We used a 
simple model to evaluate these possible effects, We do not claim that 
any credence should be placed in our very simplistic model, or that it 
is even correct. Our purpose was merely to point out that color 
fluctuations and perhaps other effects should be carefully addressed.

We further pointed out that the lack of discovery of reduced charge
fluctuations at RHIC does not imply that QGP is not formed at RHIC, since 
the overly simplified and incomplete calculations which predict this can 
easily be misleading in regard to the actual physical situation.
Furthermore the QGP signals may not have been strong enough to have been
isolated and observed by the methods used. On the other hand if reduced 
charge fluctuations are observed this would be an important observation.
However, by itself  this would not be conclusive evidence for its origin 
in a QGP, unless there is convincing other general strong supporting 
evidence, and perhaps most importantly no other viable explanation of 
the data.

\end{document}